\documentclass[preprint2]{aastex}

\slugcomment{Submitted to The Astrophysical Journal.}

\shorttitle{Dissociative Photoionization of PAH Derivatives}
\shortauthors{Rouill{\'e} et al.}

\begin{document}


\title{Dissociative Photoionization of Polycyclic Aromatic Hydrocarbon \\
    Molecules Carrying an Ethynyl Group}


\author{G. Rouill{\'e}, S. A. Krasnokutski, D. Fulvio, and C. J{\"a}ger}
\affil{Laboratory Astrophysics Group of the Max Planck Institute for Astronomy at the Friedrich Schiller University Jena, Institute of Solid State Physics, Helmholtzweg 3, D-07743 Jena, Germany}
\email{cornelia.jaeger@uni-jena.de}

\author{Th. Henning}
\affil{Max Planck Institute for Astronomy, K{\"o}nigstuhl 17, D-69117 Heidelberg, Germany}

\and

\author{G. A. Garcia, X.-F. Tang, and L. Nahon}
\affil{Synchrotron SOLEIL, L'Orme des Merisiers, Saint-Aubin BP 48, F-91192 Gif-sur-Yvette Cedex, France}




\begin{abstract}
The life cycle of the population of interstellar polycyclic aromatic hydrocarbon (PAH) molecules depends partly on the photostability of the individual species. We have studied the dissociative photoionization of two ethynyl-substituted PAH species, namely, 9-ethynylphenanthrene and 1-ethynylpyrene. Their adiabatic ionization energy and the appearance energy of fragment ions have been measured with the photoelectron photoion coincidence (PEPICO) spectroscopy technique. The adiabatic ionization energy has been found at 7.84 $\pm$ 0.02~eV for 9-ethynylphenanthrene and at 7.41 $\pm$ 0.02~eV for 1-ethynylpyrene. These values are similar to those determined for the corresponding non-substituted PAH molecules phenanthrene and pyrene. The appearance energy of the fragment ion indicative of the loss of a H atom following photoionization is also similar for either ethynyl-substituted PAH molecule and its non-substituted counterpart. The measurements are used to estimate the critical energy for the loss of a H atom by the PAH cations and the stability of ethynyl-substituted PAH molecules upon photoionization. We conclude that these PAH derivatives are as photostable as the non-substituted species in H~{\small I} regions. If present in the interstellar medium, they may play an important role in the growth of interstellar PAH molecules.
\end{abstract}



\section{Introduction}

The existence of interstellar polycyclic aromatic hydrocarbon (PAH) molecules has been inferred from the observation of infrared emission bands \citep{Leger84,Allamandola85}. It has been proposed that PAH molecules, before being blown into the interstellar medium (ISM), are formed in the envelope of carbon-rich stars at temperatures in the range 900--1700~K through the hydrogen-abstraction-C$_2$H$_2$-addition (HACA) mechanism \citep{Frenklach85,Frenklach89,Cherchneff92,Wang94,Cherchneff99}. Experiments with the pyrolysis of small hydrocarbons in the gas phase have actually shown that temperatures lower than 1800~K are required for an efficient condensation of the pyrolysis products into PAH molecules, presumably through HACA \citep{Jaeger07,Jaeger09}. In the various regions of the ISM, where temperatures are lower than 100~K, free-flying PAH molecules are thought to be formed through the destruction of carbonaceous grains in shocks \citep[see][and references therein]{Chiar13}. Additionally, despite the low temperatures, PAH species could grow through a series of chemical reactions that require little or no activation energy. Accordingly, \cite{Mebel08} have introduced the ethynyl addition mechanism (EAM), which is barrierless with regard to activation energy, as a means to form PAH molecules in cold gas-phase environments, including the conditions of the ISM.

Ethynyl-substituted PAH species appear as intermediates in the HACA mechanism and the EAM as well. Thus they may be present in the ISM, either as species ejected from stellar envelopes before their transformation into regular PAH molecules or as intermediates formed locally. The HACA mechanism has been studied as a formation process of PAH species by taking into account the conditions that prevail in stellar envelopes \citep{Frenklach89,Cherchneff92,Cherchneff99}. The EAM, however, has yet to be evaluated as an interstellar growth process, i.e., with regard to the abundance of the ethynyl radical (C$_2$H) and considering the competition with destruction mechanisms, especially photofragmentation. Relatively high amounts of C$_2$H have been found in the ISM \citep{Tucker74}, in diffuse clouds \citep{Lucas00,Gerin11,Liszt12}, in translucent clouds \citep{Turner00}, in certain dark clouds \citep{Wootten80,Ohishi92}, in circumstellar envelopes \citep{Huggins84}, in protoplanetary disks \citep{Henning10}, and in massive star-forming regions \citep{Beuther08}. On the other hand, the photostability of ethynyl-substituted PAH molecules has yet to be experimentally characterized.

\cite{Jochims99} carried out laboratory measurements on the photostability of several PAH molecules. While the set of substances included methyl-, vinyl-, and phenyl-substituted species, it did not contain ethynyl derivatives. Nevertheless, they defined a structure-dependent index of photostability that allowed them to compare the behavior of substituted PAH molecules with that of regular ones. The latter comprise exclusively fused, six-membered aromatic carbon cycles. They found that while a regular PAH would survive in H~{\small I} regions, its substituted counterparts would undergo fragmentation in the same UV radiation field. This could suggest that the ethynyl derivatives would not survive in H~{\small I} regions, in contrast with our expectation that they are photostable \citep{Rouille12,Rouille13}.

The possible presence of ethynyl-substituted PAH molecules in the ISM and the lack of data concerning their photostability called for an experimental study. We present here laboratory measurements on the dissociative photoionization of two ethynyl-substituted PAH species, 9-ethynylphenanthrene (C$_{16}$H$_{10}$) and 1-ethynylpyrene (C$_{18}$H$_{10}$). The photostability index $R$ as defined by \cite{Jochims99} can be derived from few experimental data, namely, the adiabatic ionization energy, the appearance energy of the singly dehydrogenated fragment ion, and the internal thermal energy of the parent ion. Their values have been determined in photoelectron photoion coincidence (PEPICO) spectroscopy experiments and, after deriving the value of $R$ for both ethynyl-substituted PAH species, we conclude that they are as photostable as the non-substituted molecules.

\section{Experimental details} \label{Exp}

The experiments were carried out at the beamline DESIRS\footnote{Dichro{\"i}sme Et Spectroscopie par Interaction avec le Rayonnement Synchrotron} \citep{Nahon12} of the synchrotron SOLEIL\footnote{Source Optimis{\'e}e de Lumi{\`e}re d'Energie Interm{\'e}diaire du LURE}. This beamline delivers photons with an energy in the range 5--40~eV, thus covering the VUV wavelength domain. Photons at chosen energies were used to photoionize PAH species in a molecular beam produced with the vacuum apparatus SAPHIRS\footnote{Spectroscopie d'Agr{\'e}gats PHotoIonis{\'e}s par le Rayonnement Synchrotron} \citep{RichardViard96}. The photoelectrons and the photoions -- parents and fragments -- were analyzed using the double-imaging PEPICO spectrometer DELICIOUS III \citep{Garcia13}. Briefly, it consists of a velocity-map-imaging (VMI) electron analyzer and a modified Wiley-McLaren ion imaging spectrometer operated in coincidence. The coincidence treatment provides photoelectron images that can be tagged by their corresponding ion mass and ion translational energy. From these images, photoelectron spectra can be obtained from all the masses in the molecular beam simultaneously. The energy of the VUV radiation was calibrated using the photoionization of N$_2$(X$^1\Sigma^+_g$) into N$^+_2$(B$^2\Sigma^+_u$) as a reference \citep{Innocenti13}.

We used samples of phenanthrene (C$_{14}$H$_{10}$, Aldrich, purity $\geq$99.5{\%}), 9-ethynylphenanthrene (C$_{16}$H$_{10}$, Aldrich, purity 97{\%}), pyrene (C$_{16}$H$_{10}$, Aldrich, purity 99{\%}), and 1-ethynylpyrene (C$_{18}$H$_{10}$, abcr, purity 96{\%}), as received, without undertaking a further purification.

In each experiment, the sample powder was heated in an oven placed in the source chamber of SAPHIRS, so as to increase its vapor pressure up to a useful level. The oven temperature was 383~K for phenanthrene, 493~K for pyrene, up to 443~K for 9-ethynylphenanthrene (413~K for energy-dependent breakdown measurements and 443~K for photoelectron spectroscopy measurements; see below in this Section), and 403~K for 1-ethynylpyrene. In order to prevent its obstruction by condensing sample vapor, the nozzle was heated to a temperature 10~K higher than that of the oven. The phenanthrene and pyrene powders were poured in a stainless-steel boat that was inserted into the oven in a free fit. A sufficient amount of the 9-ethynylphenanthrene sample was transferred into the gas phase by following the same procedure, although most of the substance polymerized and remained in the oven. In the case of 1-ethynylpyrene, however, it was necessary to finely ground the sample in a mortar; the fine powder was dispersed in a ball of ultrafine glass wool, which was placed in the boat in the oven. Increasing the surface area of the sample improved the transfer of molecules into the gas phase, allowing us to work with a lower oven temperature and for a longer time with the same sample load.

In the oven, the PAH molecules diffused in an atmosphere of helium gas (Air Liquide, purity $\geq$99.999{\%}) provided with a backing pressure of 0.5 to 1~bar. The mixed vapor leaked out of the oven through a nozzle 70~$\mu$m in diameter, producing a continuous, supersonic jet in the source chamber. This chamber is the first of the three differentially pumped vacuum chambers of SAPHIRS, which communicate by means of skimmers 1~mm in diameter. The jet and the skimmers separating the chambers were aligned so as to produce a molecular beam of He atoms and PAH molecules. In the third chamber, the molecular beam crossed the synchrotron radiation beam in the ionization volume of the PEPICO spectrometer.

Adiabatic ionization energies were obtained from the mass-selected photoelectron images recorded at a fixed photon energy of 9~eV. The proximity of this energy to the ionization threshold of the molecules led to the production of slow electrons and the possibility to lower the extraction field to 90 V~cm$^{-1}$ with the consequent gain in kinetic energy resolution \citep{Garcia13}. The corresponding mass-selected photoelectron spectra were obtained from the images applying the pBasex algorithm for Abel inversion \citep{Garcia04}. A gas filter upstream the monochromator was filled with Kr to effectively cut off the high harmonics of the undulator \citep{Mercier00}. For these measurements, the photoelectrons and ions were counted during 6483~s for phenanthrene, 3627~s for 9-ethynylphenanthrene, and 18\,034~s for 1-ethynylpyrene.

The appearance energy of the ion fragments indicative of dissociative photoionization was measured by varying the photon energy from 15 to 20~eV with a step of 0.1~eV. The gas filter was not used for these measurements as the high harmonics were not transmitted by the optics. At each energy the ion signal was integrated for 180~s for all masses simultaneously. In the case of 1-ethynylpyrene, two sets of measurements were averaged to improve the signal-to-noise ratio. In this kind of experiments, the extraction field of the spectrometer was set to provide full transmission for all the electrons and ions present in the sample. All the data were corrected for false coincidences and normalized for the incident photon flux as measured by an IRD AXUV100 photodiode.

\section{Results} \label{Res}

\subsection{Adiabatic ionization energies} \label{Res:AIEs}

\begin{figure}
\epsscale{1.00}
\plotone{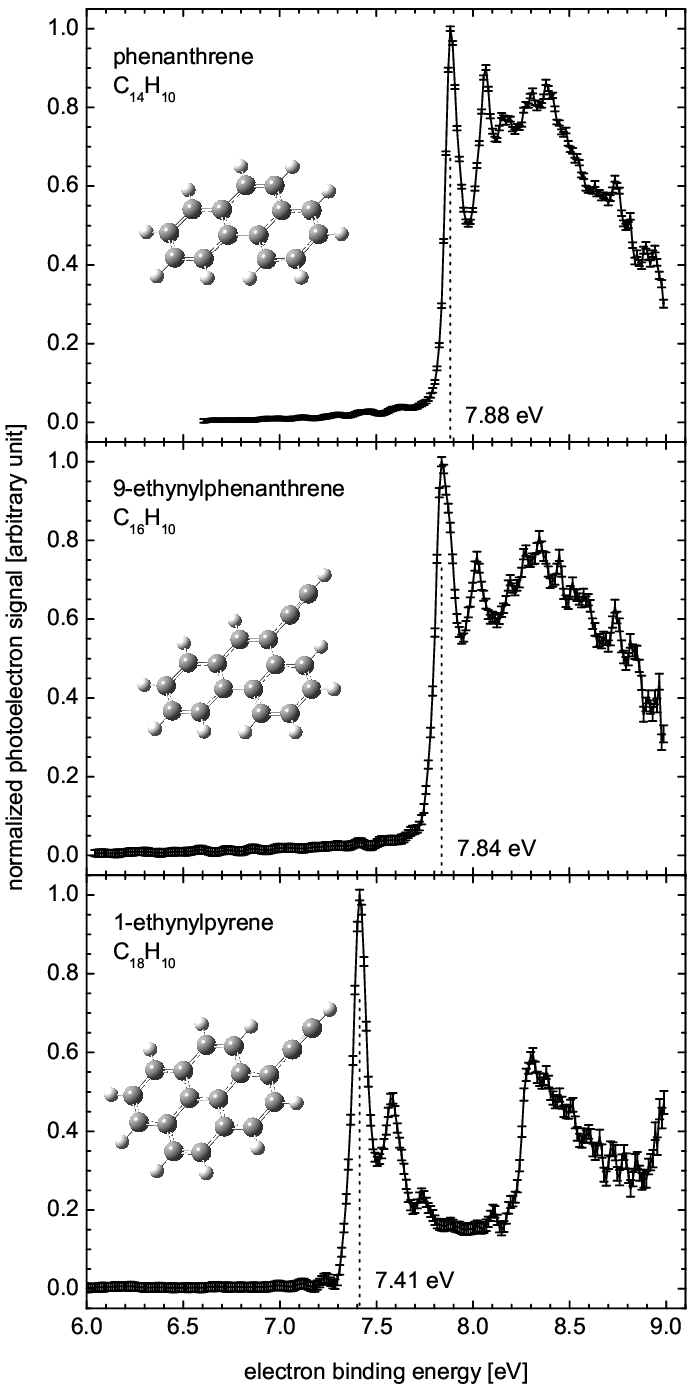}
\caption{Photoelectron spectra measured with photons of 9 eV in energy. The vertical dotted lines indicate the adiabatic ionization energies. The error bars are obtained assuming a Poisson distribution on the image pixel intensities and then propagating the incertitudes across all algebra operations involving the Abel inversion.\label{fig1}}
\end{figure}

The adiabatic ionization energies $E_{\mathrm{i}}$ of phenanthrene, 9-ethynylphenanthrene, and 1-ethynylpyrene have been determined by analyzing photoelectron spectra. These spectra were extracted from the electron images obtained in the PEPICO measurements carried out with a fixed photon energy equal to 9~eV. The spectra are shown in Figure~\ref{fig1} and the experimental ionization energies are given in Table~\ref{tbl-1}. As detailed below, the ionization energy of pyrene was taken from the literature.

We have measured the adiabatic ionization energy of phenanthrene to demonstrate the calibration of the spectrometer as described in Section~\ref{Exp}. We found this energy at 7.88 $\pm$ 0.02~eV, in agreement with the various values reported in the literature, which were determined with different techniques: 7.86 $\pm$ 0.10~eV \citep[][photoelectron spectroscopy]{Boschi72b}, 7.90 $\pm$ 0.01--0.03~eV \citep[][photoion mass spectrometry, fully deuterated phenanthrene]{Jochims94}, 7.903 $\pm$ 0.005~eV \citep[][resonant two-photon ionization]{Thantu93}, 7.8914~eV \citep[][two-laser ionization mass spectrometry]{Hager88}, and 7.87 $\pm$ 0.02~eV \cite[][time-resolved photoionization mass spectrometry, TPIMS]{Gotkis93b}.

As the adiabatic ionization energy we have obtained for phenanthrene is in agreement with the values reported in the literature, we have adopted for the ionization energy of pyrene the most recent value of 7.415 $\pm$ 0.010~eV that was determined in a PEPICO experiment \citep{Mayer11}. It is consistent with other reports: 7.41~eV \citep[][photoelectron spectroscopy]{Boschi72a}, 7.45 $\pm$ 0.01--0.03~eV \citep[][photoion mass spectrometry]{Jochims94}, 7.4251 $\pm$ 0.0009~eV \citep[][coupled resonance-enhanced multiphoton ionization and zero kinetic energy photoelectron spectroscopies, REMPI-ZEKE, value obtained by conversion of 59\,888 $\pm$ 7~cm$^{-1}$]{Zhang10}, and 7.4256~eV \cite[][two-laser ionization mass spectrometry]{Hager88}.

For 9-ethynylphenanthrene and 1-ethynylpyrene, the adiabatic ionization energies were found at 7.84 $\pm$ 0.02~eV and 7.41 $\pm$ 0.02~eV, respectively. Thus the ionization energy of the ethynyl-substituted PAH molecules is equal to that of the corresponding non-substituted species within the experimental error bars. This is in contrast to the lower ionization potentials reported by \cite{Jochims99} for methyl-, vinyl-, and phenyl-substituted PAH species with respect to the non-substituted molecules. They measured differences not smaller than 0.15 eV.

In addition to ionization energies, the photoelectron spectra of Figure~\ref{fig1} reveal low-lying electronic states of the cations. The peak at 8.07~eV in the spectrum of phenanthrene likely marks the origin of a transition to such a state. This ionization was measured at 8.10~eV by \cite{Boschi72b} and semi-empirical calculations actually predicted the first excited doublet state of the cation to lie about 0.5~eV above its ground state \citep{Khan92,Parisel92}. Since the photoelectron spectra of 9-ethynylphenanthrene and phenanthrene are essentially similar, we conclude that the cation of the derivative possesses an equivalent state. The spectrum measured with 1-ethynylpyrene clearly shows a feature with an origin at 8.3~eV, indicating that the cation has an electronic doublet state 0.9~eV above its ground state. The pyrene cation also exhibits such a low-lying electronic state \citep{Boschi72a}.

\subsection{Appearance energies} \label{Res:AEs}

Breakdown graphs describe the fraction of all detected ions, parent and fragments, as a function of the energy of the incident photons.

\begin{figure}
\epsscale{1.00}
\plotone{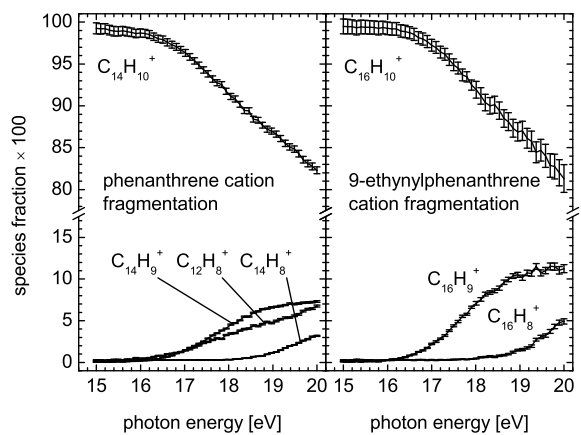}
\caption{Breakdown graphs of the dissociative photoionization of phenanthrene and 9-ethynylphenanthrene. The error bars result from a Poisson distribution on the integrated ion signal and standard-error propagation formulae.\label{fig2}}
\end{figure}

In Figure~\ref{fig2}, the breakdown graph for the dissociative photoionization of phenanthrene (C$_{14}$H$_{10}$) between 15 and 20~eV shows the appearance of three fragment ions revealing the loss by the parent ion of one H atom (C$_{14}$H$_9^+$), of two H atoms (C$_{14}$H$_8^+$), and of two C and two H atoms (C$_{12}$H$_8^+$). The same figure reports the breakdown graph obtained in the experiment on 9-ethynylphenanthrene (C$_{16}$H$_{10}$), during which only the two fragment ions indicating the loss of one H atom (C$_{16}$H$_9^+$) and the loss of two H atoms (C$_{16}$H$_8^+$) were detected.

\begin{figure}
\epsscale{1.00}
\plotone{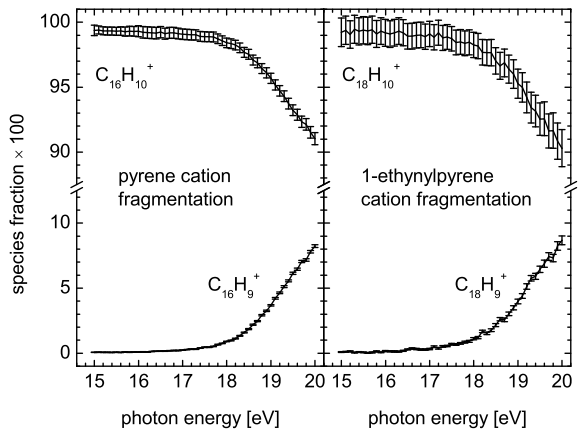}
\caption{Breakdown graphs of the dissociative photoionization of pyrene and 1-ethynylpyrene. The error bars are described in the caption of Figure~\ref{fig2}.\label{fig3}}
\end{figure}

The breakdown graphs derived from the measurements on pyrene (C$_{16}$H$_{10}$) and 1-ethynylpyrene (C$_{18}$H$_{10}$) are displayed in Figure~\ref{fig3}. In either case, only the fragment ion corresponding to the loss of a single H atom was detected in the 15--20~eV interval. This is consistent with the breakdown graph in the recent PEPICO study of the dissociative photoionization of pyrene by \cite{West14b}. In that study, the next fragment ion, which corresponds to the loss of two H atoms, appeared at $\approx$20~eV, the upper limit of the energy domain we have scanned.

The appearance energy $E_{\mathrm{a}}$ of a fragment ion, i.e., the lowest photon energy for which it is observed, is an effective value since it depends on the sensitivity of the experiment. Indeed, a slower fragmentation rate requires a longer time interval to produce a number of fragment ions that is sufficient for a sensible counting. Thus the time scale of the measurement affects the determination of the appearance energy -- an effect known as the kinetic shift -- as illustrated with TPIMS experiments \citep[see, for instance,][]{Lifshitz91}. With the photoionization of phenanthrene in a trap, the appearance energy of the singly dehydrogenated ion $E_{\mathrm{a},-\mathrm{H}}$ was found to be 14.9 $\pm$ 0.1~eV when measured 24~$\mu$s after ionization and 14.0 $\pm$ 0.2~eV after 5~ms \citep{Ling98}. Similarly, TPIMS experiments on pyrene yielded 16.2 $\pm$ 0.2~eV at 24~$\mu$s and 15.2 $\pm$ 0.2~eV at 5~ms for the appearance energy of the singly dehydrogenated ion \citep{Ling95}.

\begin{figure}
\epsscale{1.00}
\plotone{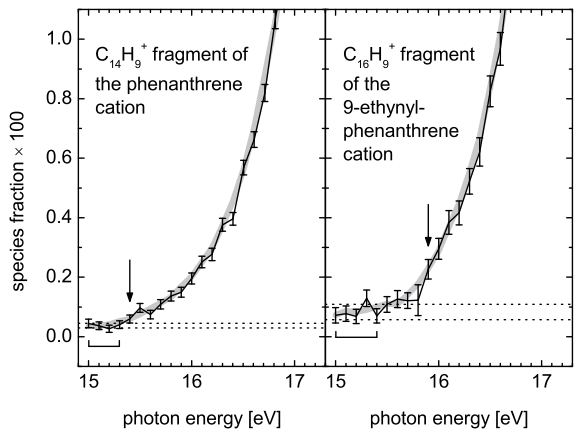}
\caption{Onset of H loss by the phenanthrene and 9-ethynylphenanthrene cations. The error bars are described in the caption of Figure~\ref{fig2}. The horizontal dotted lines frame the background signal, which is centered at the mean value of the points in the interval defined with the horizontal bar and extends on both side by one sample standard deviation. The arrow indicates the first of two consecutive points for which the signal and its error bar are above the background, the energy corresponding to this point being taken as the appearance energy of the fragment ion. The thick gray line represents a centered five-point moving average of the measurements.\label{fig4}}
\end{figure}

\begin{figure}
\epsscale{1.00}
\plotone{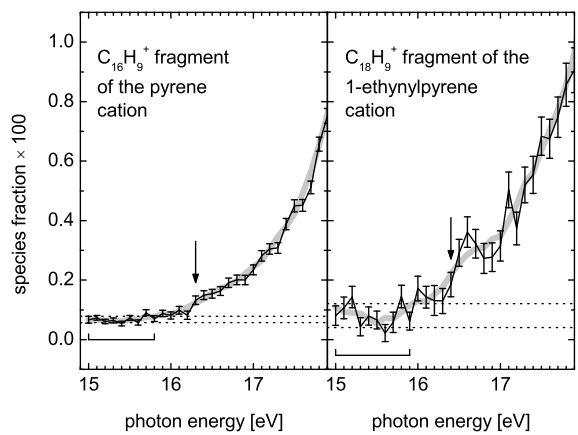}
\caption{Onset of H loss by the pyrene and 1-ethynylpyrene cations. The error bars are described in the caption of Figure~\ref{fig2}. The other items are defined in the caption of Figure~\ref{fig4}.\label{fig5}}
\end{figure}

In our PEPICO spectroscopy experiment, the detected photofragmentations occurred within microseconds, the time interval necessary for the ionized molecules to leave the acceleration zone. Figures~\ref{fig4} and \ref{fig5} are expanded views of the breakdown graphs showing the onset of H loss by the cations we have studied. Considering the microsecond time scale of our measurements, the appearance energies determined for phenanthrene and pyrene, respectively 15.4 $\pm$ 0.1~eV and 16.3 $\pm$ 0.1~eV, are consistent with those derived previously in TPIMS experiments and reminded above. Moreover, in the case of pyrene, the value is the same as that reported by \cite{Jochims94}, i.e., 16.30 $\pm$ 0.15~eV, showing that the two setups have equal sensitivities in terms of fragmentation rate.

Comparing the dissociative photoionization of each ethynyl-substituted PAH molecule with the corresponding non-substituted species, the appearance energies of the singly dehydrogenated ions stand out by their proximity, with a difference of 0.5~eV for the phenanthrenes and 0.1~eV for the pyrenes. In contrast, \cite{Jochims99} had observed in their experiments that the minimum photon energy causing the loss of a H atom by a photoionized PAH molecule was at least 2~eV lower when the molecule carried a methyl, a vinyl, or a phenyl group. Prior to the latter study, \cite{Gotkis93a} and \cite{Gotkis93b} had found in TPIMS experiments that the minimum photon energy for H atom loss upon photoionization was also at least 2~eV lower in 1- and 2-methylnaphthalene than in naphthalene, for any time scale. 

The different behaviors of the diverse PAH derivatives with respect to dissociation are explained by the various degrees of conjugation of their $\pi$ electronic system \citep{Jochims99}. Presently, the CC triple bond of an ethynyl side group yields a stronger conjugation of the $\pi$ electronic orbitals over the whole molecule in comparison with the CC double bond of a vinyl group, the aromatic ring of a phenyl group, and the saturated C atom of a methyl group. This is valid for the neutral species and the cations as well. As a result, ethynyl-substituted PAH molecules exhibit a strengthened structure.

With regard to the origin of the H atom lost by the ethynyl-substituted PAH molecules upon photoionization, we propose that it is one of the H atoms attached to the aromatic rings rather than the H atom of the ethynyl side chain. This proposition is first suggested by the conjunction of similar ionization energies and similar appearance energies of the singly dehydrogenated ions for the regular and ethynyl-substituted PAH molecules. The proposition in also supported by the fact that the dissociation energy of a CH bond is lower for PAH molecules \citep[469 $\pm$ 6~kJ mol$^{-1}$ for naphthalene; see][]{Reed00} than for acetylene \citep[549 $\pm$ 4~kJ mol$^{-1}$,][]{Ervin90} or diacetylene \citep[539 $\pm$ 12~kJ mol$^{-1}$,][]{Shi00}. This difference holds for the cations \citep[4.48~eV for the naphthalene cation and 5.92~eV for the acetylene cation; see, respectively,][]{Ho95,vanderMeij88}.

Interestingly, the fragmentation behavior of the present ethynyl-susbtituted PAH cations differs from that of ethynylbenzene (also called phenylacetylene). In an imaging PEPICO experiment, \cite{West14a} did not found any H loss channel while they measured the loss of a C$_2$H$_2$ unit. Our observations of the dissociative photoionization of the ethynyl-susbtituted PAH species are opposite since the loss of a H atom is clearly detected while that of a C$_2$H$_2$ unit is not.

\subsection{Photostability} \label{Res:PhotoStab}

\cite{Jochims99} designed the photostability index $R$ in the framework of the Rice-Ramsperger-Kassel theory for unimolecular reaction rates \citep[RRK theory;][]{Rice27,Kassel28}. The index was introduced to examine the behavior of PAH derivatives with regard to fragmentation relative to that of the regular species. It compares $E_{\mathrm{c},-\mathrm{H}}$, the critical internal energy for the loss of a H atom by a PAH cation, with $E_{-\mathrm{H}}\left(10^2\right)$, the internal energy causing the loss of a H atom by a regular PAH cation at the rate of 10$^2$ s$^{-1}$. While $E_{\mathrm{c},-\mathrm{H}}$ is derived from measurements, $E_{-\mathrm{H}}\left(10^2\right)$ is the result of a calculation using a model of the fragmentation of regular PAH cations. The rate of 10$^2$ s$^{-1}$ represents the limit under which fragmentation is not efficient as it is slower than relaxation through emission of IR photons. Thus
\begin{equation} \label{eq-R}
R = \frac{E_{\mathrm{c},-\mathrm{H}}}{E_{-\mathrm{H}}\left(10^2\right)} .
\end{equation}
The value of $E_{-\mathrm{H}}\left(10^2\right)$ is computed using the RRK theory. The reaction rate for the loss of a H atom $k_{-\mathrm{H}}$ is then defined with
\begin{equation}
k_{-\mathrm{H}} = \nu \left( 1 - \frac{E_{0,-\mathrm{H}}}{E} \right)^{s - 1} ,
\end{equation}
where $E$ is the internal energy of the PAH ion, $s$ is the number of its vibrational modes, $E_{0,-\mathrm{H}}$ = 2.8~eV is the average activation energy for the loss of a H atom by a PAH species, and $\nu$ = 10$^{16}$~s$^{-1}$ is a frequency factor \citep{Jochims94}. It follows that $E_{-\mathrm{H}}\left(k_{-\mathrm{H}}\right)$, the internal energy corresponding to this rate, is obtained with
\begin{equation}
E_{-\mathrm{H}}\left(k_{-\mathrm{H}}\right) = \frac{E_{0,-\mathrm{H}}}{1 - \left( k_{-\mathrm{H}} / \nu \right)^{1 / \left( s - 1 \right) } } .
\end{equation}
The critical internal energy for fragmentation is derived from the experimental appearance energy of the singly dehydrogenated ion $E_{\mathrm{a},-\mathrm{H}}$, corrected for the kinetic shift $\Delta E$, from which is subtracted the ionization energy $E_{\mathrm{i}}$. Moreover, the thermal contribution $E_{\mathrm{t}}$ to the internal energy has to be taken into account. Thus
\begin{equation} \label{eq-Ec-H}
E_{\mathrm{c},-\mathrm{H}} = E_{\mathrm{a},-\mathrm{H}} - \Delta E - E_{\mathrm{i}} + E_{\mathrm{t}} .
\end{equation}
The kinetic shift $\Delta E$ is evaluated according to
\begin{equation}
\Delta E = E_{-\mathrm{H}}\left(10^4\right) - E_{-\mathrm{H}}\left(10^2\right),
\end{equation}
i.e., the difference between the internal energy corresponding to the slowest fragmentation rate that can be detected in our experiment and the internal energy giving the slowest rate for a fragmentation faster than a radiative deexcitation. We take the value 10$^4$~s$^{-1}$ determined by \cite{Jochims94} for the slowest fragmentation rate that can be detected because the appearance energy we have measured in the case of pyrene is the same as theirs. The computed kinetic shifts are given in Table~\ref{tbl-1}. They are consistent with the results of the TPIMS experiments that showed a kinetic shift of $\approx$1~eV for dissociations measured at the microsecond time scale, taking the measurements at the millisecond time scale as references for infinite time \citep{Ling95,Ling98}.

Like \cite{Jochims94}, we take the average internal thermal energy $\left\langle E_{\mathrm{t}} \right\rangle$ to represent the thermal contribution $E_{\mathrm{t}}$ to the internal energy. The internal thermal energy is assumed to be purely vibrational by nature. Consequently the average internal thermal energy can be computed from the vibrational partition function. Thus
\begin{equation} \label{eq-Eth}
\left\langle E_{\mathrm{t}} \right\rangle = \sum^{s}_{i=1} \frac{h\nu_i}{\mathrm{exp} \left( h\nu_i / kT \right)- 1} ,
\end{equation}
where $h$ is Planck's constant, $\nu_i$ is the frequency of mode number $i$, $k$ is Boltzmann's constant, and $T$ is the vibrational temperature. For the calculation, we use scaled theoretical frequencies. The vibrational modes of the neutral ethynyl derivatives were previously computed in an application of the density functional theory \citep{Rouille12}. Collective scaling factors for frequencies below and above 2000~cm$^{-1}$ were derived to fit the theoretical harmonic frequencies to band positions observed for the substances imbedded in CsI pellets \citep{Rouille12}. We have calculated the vibrational modes of neutral phenanthrene and neutral pyrene at the same level of theory using the same Gaussian 09 software \citep{Gaussian09}. Frequency scaling factors were derived for these two regular PAH molecules simultaneously using the spectra measured in argon matrix by \cite{Hudgins98} as references. The scaling factor is 0.979 for all frequencies except those of the CH stretching modes, for which it is 0.964.

The vibrational temperature of the molecules would be comprised between their translational temperature and the temperature of the oven, the latter being given in Section~\ref{Exp}. The translational temperature has been derived for each molecular species from its velocity along the molecular beam coordinates, which has been determined by analyzing the ion image. Each translational temperature has been rounded to the tens of K because unknown factors such as the exact composition of the molecular beam give uncertainties that are of this order. Thus a temperature interval has been defined for each species using its translational temperature as the lower limit and the oven temperature as the upper one. All temperatures and the internal thermal energies derived from them are reported in Table~\ref{tbl-1}. Table~\ref{tbl-1} also contains the resulting critical energies $E_{\mathrm{c},-\mathrm{H}}$ and photostability indexes $R$. Although the translational temperature differs largely from the oven temperature, the presently modest contribution of $E_{\mathrm{t}}$ (or $\left\langle E_{\mathrm{t}} \right\rangle$) to $E_{\mathrm{c},-\mathrm{H}}$ results in a small dependence of $R$ on $T$.

The first result of this study is the consistency of the $R$ values presently derived for phenanthrene and pyrene, (0.929--0.968) $\pm$ 0.017 and (1.057--1.114) $\pm$ 0.015, respectively, with those given by \cite{Jochims99}, 0.89 and 1.05, respectively. One may note that the latter values, the accuracy of which was not given, lie at the lower end of the intervals we have obtained through our analysis. This may indicate that the vibrational temperature of the molecules in the helium beam is closer to their translational temperature than to the oven temperature. Nevertheless, the values are close to 1, which is the ideal index value for a regular PAH. The simplicity of the model used to determine $E_{-\mathrm{H}}\left(k_{-\mathrm{H}}\right)$ leads to the observed deviations. The second result is the similarity between the index values obtained for an ethynyl derivative and for the corresponding regular PAH molecule. From this similarity we conclude that the ethynyl derivatives exhibit the same photostability as the regular species.

In H~{\small I} regions, molecules may interact with photons carrying a maximum energy of 13.6~eV. The stability of a neutral PAH molecule upon photoionization is evaluated by comparing the critical energy for the loss of a H atom by a cation, $E_{\mathrm{c},-\mathrm{H}}$, with the maximum internal energy that can be transferred to this cation, $E^{\ast}$, where $E^{\ast} = 13.6 - E_{\mathrm{i}}$ \citep{Jochims99}. As shown in Table~\ref{tbl-1}, $E_{\mathrm{c},-\mathrm{H}}$ is larger than $E^{\ast}$ for the four substances we have studied. As a consequence, they would be photostable in H~{\small I} regions. As seen at the beginning of the section, the stability index used throughout relies on the H loss occurring within a hot ground state, after internal conversion. Non-statistical processes such as direct dissociation from an excited repulsive state might happen although previous works have favoured the statistical path considered here \citep[see][]{Jochims94,Allain96a,LePage01}.

\begin{figure}
\epsscale{1.00}
\plotone{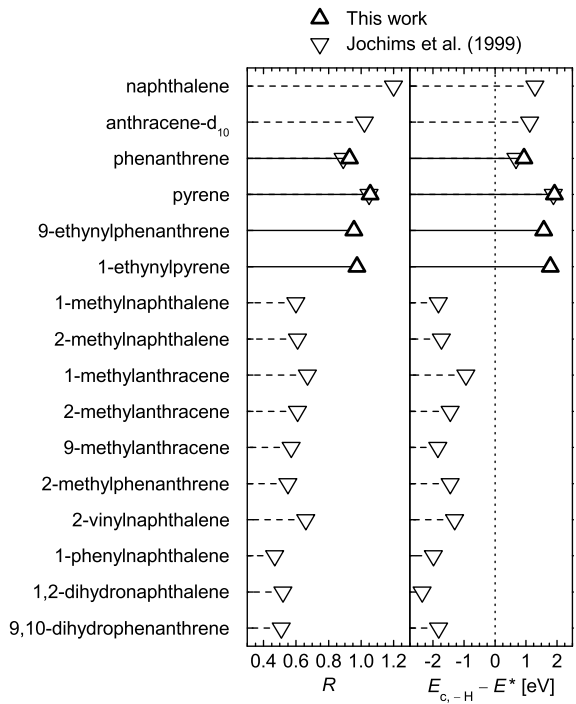}
\caption{Comparison of the $R$ values (left) and of the $E_{\mathrm{c},-\mathrm{H}} - E^{\ast}$ differences (right) for several PAH molecules. The vertical dotted line in the right panel indicates the limit above which PAH molecules are stable in the UV radiation field of H~{\small I} regions.\label{fig6}}
\end{figure}

Figure~\ref{fig6} presents for comparison some of the photostability index values determined by \cite{Jochims99} for various regular PAH molecules and derivatives along with those we have derived by considering that the internal thermal energy of the molecules reflected their translational temperature. It also reports the $E_{\mathrm{c},-\mathrm{H}} - E^{\ast}$ differences. The figure shows that the ethynyl derivatives of PAH molecules are as photostable as the corresponding regular species while the other derivatives are clearly less photostable and are actually expected to fragment upon photoionization in H~{\small I} regions \citep{Jochims99}.

\section{Discussion} \label{Disc}

In a theoretical study, \cite{Allain96a} proposed a criterion for the survival of interstellar PAH molecules with respect to photofragmentation. Defining the molecules by their carbon content, the survival requires that the rate for the loss of a C$_2$H$_2$ group be not faster than the rate for the addition of carbon atoms, either neutral or ionized. After computing and comparing the corresponding loss and addition rates for PAH molecules, they reached the conclusion that only those containing more than 50 C atoms survive in the ISM. Of relevance to the present work, after \cite{Allain96a} noted the production of ethynyl derivatives through the HACA mechanism, hence their possible presence in the ISM, they computed the loss rate of a C$_2$H$_2$ group for ethynyl-substituted PAH molecules. The values they reported did not indicate a substantial difference with regular PAH species, which is consistent with our experimental breakdown graphs for phenanthrene and ethynylphenanthrene. These graphs would even suggest a faster rate for the loss of a C$_2$H$_2$ group by non-substituted species since the corresponding fragment was not observed for the ethynyl derivative. This single instance of comparison, however, may not be representative. We can still state, taking the survival criterion defined by \cite{Allain96a}, that ethynyl-substituted PAH molecules containing more than 50 C atoms survive in the ISM as far as photofragmentation is concerned.

Rather than defining the destruction of a PAH molecule by the loss of C atoms, one may regard the loss of a H atom by the PAH molecule as the decisive step toward further destruction. Indeed, the loss of a H atom is the unimolecular dissociation mechanism that requires the least internal energy \citep{Jochims99} and the dehydrogenation results in the weakening of the molecular structure \citep{Allain96b,Jochims99}. Using Jochims et al.'s photostability index $R$, which is built on the loss of a H atom, we have determined that 9-ethynylphenanthrene and 1-ethynylpyrene were as photostable as phenanthrene and pyrene, respectively. These four molecules were also found to be stable upon photoionization in H~{\small I} regions, unlike every PAH molecule carrying a methyl, vinyl, or phenyl group studied by \cite{Jochims99}. Note that the same study reported the dihydrogenated PAH species as undergoing dissociative photoionization in H~{\small I} regions, too, as shown in Figure~\ref{fig6}. As mentioned in Section~\ref{Res:AEs}, the degree of conjugation of the $\pi$ orbitals affects the stability of the molecular structures. Concerning dihydrogenated derivatives, the addition of H atoms lowers the conjugation by creating adjacent single CC bonds, weakening the structure. As a conclusion, by examining the photostability of PAH molecules using $R$, we find that the ethynyl derivatives of regular PAH molecules that survive in the UV radiation field of H~{\small I} regions would also survive in this environment, i.e., in most of the ISM. In terms of size, \cite{Jochims94} found that survival required of the regular PAH molecules that they contained at least 30 C atoms. While this limit was derived from observing the fragmentation of the cations, it also applies to the neutrals, as stated at the end of Section~\ref{Res:PhotoStab}. The same requirement with respect to size can be extended to the ethynyl derivatives in view of their similar fragmentation pattern.

As photostable interstellar molecules, ethynyl-substituted PAH species would survive long enough to react with other interstellar substances, e.g., C$_2$H (see Introduction). In the EAM scheme, a PAH molecule would undergo a series of reactions with C$_2$H leading to the formation of new aromatic cycles and, as a consequence, the growth of the molecule \citep{Mebel08,Jones10}. As an experimental argument favorable to the EAM, it was found that C$_2$H reacted with benzene (C$_6$H$_6$) in the gas phase in a barrierless mechanism that produces ethynylbenzene (phenylacetylene, C$_6$H$_5$C$_2$H) plus a H atom \citep{Goulay06a,Jones11}. Note that C$_2$H reacts also with acetylene (C$_2$H$_2$), giving diacetylene (C$_4$H$_2$) and a H atom. This process is also free of energy barrier at the entrance channel as shown by \cite{Lee00} with another experiment in the gas phase. Thus C$_2$H may react with ethynyl-substituted PAH molecules under interstellar conditions, at an aromatic cycle or at the side chain, as suggested by the different paths described by \cite{Mebel08} for the reaction of C$_2$H with ethynylbenzene.

The C$_2$H radical, however, is not the only abundant interstellar species that may be involved in the growth of PAH molecules. Neutral atomic carbon (C~{\small I}) is also abundant in the diffuse ISM \citep{Jenkins11} and in denser regions \citep[see, for instance,][]{Phillips81,Schilke95,Stark96,Beuther14}. This induced \cite{Allain96a} to view the addition of carbon atoms as the growth mechanism that would compete effectively with the destruction of interstellar PAH molecules. They assumed that C atoms would react with PAH molecules under the conditions of the ISM and that the reactions would contribute to the growth of the molecules rather than to their destruction. It has actually been demonstrated experimentally that C atoms react with benzene in a barrierless addition that gives a radical with a seven-membered carbon ring plus a H atom \citep{Hahndorf02}. Therefore barrierless reactions between C atoms and PAH molecules can be considered. Such reactions, however, have yet to be studied in the laboratory. Finally, experiments have shown that C atoms react with C$_2$H$_2$ molecules without energy barrier, where C + C$_2$H$_2$ $\rightarrow$ C$_3$H + H \citep{Kaiser96,Chastaing01}. Thus it is worthwhile to examine whether carbon atoms react with ethynyl-substituted PAH species, either at the polycyclic aromatic moiety or at the ethynyl side group.

In addition to reactions involving C$_2$H and C~{\small I}, one could also think of a possible role of reactions with methylidyne (the CH radical, also carbyne) in the growth of PAH molecules. Methylidyne is abundant in the ISM \citep[see, for instance,][]{Danks84,Gerin10} and is known to react at low temperatures with C$_2$H$_2$ \citep{Canosa97,Maksyutenko11} as well as with anthracene \citep{Goulay06b}.

Thus several types of neutral-neutral reactions may participate to the chemical network that constitutes the growth process of PAH species in the ISM. When considering barrierless reactions only, the growth rate depends essentially on the frequency of collisions between the reactants, i.e., on the abundance of the latter and on the temperature of the environment. Any environment, however, has a limited lifetime even if it is of the order of 10$^7$~yr as in the case of molecular clouds \citep[see, for instance,][]{Larson81}. It has yet to be examined whether the abundance of the relevant atomic and molecular species in the diverse interstellar environments allows the growth of PAH molecules within the lifetime of these environments. Moreover, we have only considered neutral-neutral chemical reactions that may lead to the addition of C atoms to PAH molecules, thus contributing to their growth. The complete chemistry of PAH species in the ISM, however, comprises other mechanisms, e.g., ion-neutral reactions, electron attachment, double ionization, and multiphoton processes, which have been evaluated with various models \citep{LePage01,LePage03,Montillaud13}. In support to double ionizazion, a recent experimental study by \cite{Zhen15} concluded that, in the ISM, photoionization prevailed over fragmentation in PAH species containing about 50 C atoms and more, leading to the formation of doubly and even triply charged cations. As a consequence, these large species were found to be photostable in interstellar conditions, in agreement with previous theoretical studies \citep{Allain96a,Montillaud13}.

Beside their place in the life cycle of interstellar PAH molecules, ethynyl derivatives are interesting for their spectroscopic properties that suggest a link with the still unidentified carriers of the diffuse interstellar bands \citep[DIBs; see, for instance,][]{Herbig95}. In previous works, we measured the UV/vis absorption bands of C$_2$H- and C$_4$H-substituted species isolated in Ne matrices. The spectra led us to propose photostable substituted PAH molecules, or similar species, as valid candidates for the carriers of the DIBs \citep{Rouille12,Rouille13}. It was indeed observed that the addition of an acetylenic side chain to a regular PAH lowers the peak intensity and broadens the profile of the strong absorption bands that these species typically show at UV wavelengths (the $\beta$-bands), where DIBs are not observed \citep{Clayton03,Gredel11,Salama11,Bhatt15}. Consequently regular PAH molecules have become less plausible as potential DIB carriers. Nonetheless, they may contribute with other PAH species to the galactic extinction curve as a rich mixture of PAH substances gives a smooth absorption spectrum \citep{Steglich11,Steglich12}. The population of a DIB carrier would be larger than that of any other PAH substance due to a favorable photoselective mechanism. The present measurements on the dissociative photoionization of ethynyl-substituted PAH molecules reveal that such species can survive in the UV radiation field of H~{\small I} regions until they react with another substance. As the time interval between the formation and the chemical destruction of the molecules would be long enough to allow them to absorb photons and relax numerous times, the molecules would leave their mark in the absorption spectrum of the ISM. The corresponding bands would peak out of the galactic extinction curve provided that large enough populations were formed. Thus the present experimental results give support to our proposal.

\section{Conclusions}

The study of the dissociative photoionization of two ethynyl-substituted PAH species has yielded their photoionization energy and the appearance energy of some photofragments. While the adiabatic ionization energies are not affected by the addition of the ethynyl chain, the appearance energies for the H-loss channel are slightly increased in the ethynyl derivatives, especially in the case of phenanthrene where an inhibition of the C$_2$H$_2$-loss channel is also seen.

We have determined the photostability index $R$ of each ethynyl derivative and found that it is similar to the value derived for the corresponding non-substituted PAH molecule. Thus these derivatives are as photostable as the regular PAH species. Consistently, having found through the determination of $E_{\mathrm{c},-\mathrm{H}} - E^{\ast}$ that the regular PAH molecules phenanthrene and pyrene are stable upon photoionization by the photons typical of H~{\small I} regions, we have found that their ethynyl derivatives would also survive this process.

Our measurements on phenanthrene and its ethynyl derivative suggest that the presence of the side chain does not increase the unimolecular reaction rate for the loss of a C$_2$H$_2$ unit. This supports the theoretical results of \cite{Allain96a} who did not found a significant variation of this rate when examining ethynyl-substituted and non-substituted PAH species.

Thus ethynyl-substituted PAH molecules appear to be as photostable as the non-substituted species whether one considers the photostability index $R$ defined by \cite{Jochims99} or the survival criterion proposed by \cite{Allain96a}. Size matters for survival, however. As a consequence, a conservative conclusion is that ethynyl-substituted PAH molecules containing at least 50 C atoms would survive the interstellar UV radiation field. Whether they are injected into the ISM or formed there, the molecules would grow through EAM at a rate that depends on the abundance of C$_2$H. Reactions with  C~{\small I} and CH may also play a role in the growth of the PAH species and ought to be studied experimentally. We have mentioned neutral-neutral reactions because experimental results are reported in the literature, yet processes involving reactions at low temperature between neutral PAH molecules and ions such as the abundant C~{\small II} should be investigated as well.

Finally, the present results support our proposal that DIB carriers could be PAH molecules whose photostability is enhanced by the presence of side groups.

\acknowledgments

This work was carried out within the framework of a cooperation between the Max-Planck-Institut f{\"u}r Astronomie, Heidelberg, and the
Friedrich-Schiller-Universit{\"a}t Jena. We acknowledge SOLEIL for provision of synchrotron radiation facilities under project No. 20140337 and we would like to thank Jean-Fran{\c c}ois Gil for assistance in using beamline DESIRS. The financial support of the CALIPSO Consortium for Transnational Access funded by the European Commission under its Seventh Framework Program is gratefully acknowledged by GR and SK.

\clearpage

\clearpage

\clearpage

\clearpage

\clearpage

\clearpage

\clearpage

\begin{deluxetable}{lllllllll}
\tabletypesize{\scriptsize}
\tablecaption{Photostability Index of PAH Molecules and Related Parameters\label{tbl-1}}
\tablewidth{0pt}
\tablehead{
\colhead{} & \colhead{$E_{\mathrm{i}}$} & \colhead{$E^{\ast}$} & \colhead{$E_{\mathrm{a},-\mathrm{H}}$} & \colhead{$\Delta E$} & \colhead{$T$\tablenotemark{a}} & \colhead{$\left\langle E_{\mathrm{t}} \right\rangle$} & \colhead{$E_{\mathrm{c},-\mathrm{H}}$} & \colhead{} \\
\colhead{Species} & \colhead{(eV)} & \colhead{(eV)} & \colhead{(eV)} & \colhead{(eV)} & \colhead{(K)} & \colhead{(eV)} & \colhead{(eV)} & \colhead{$R$}
}
\startdata
phenanthrene & 7.88 $\pm$ 0.02 & 5.72 & 15.4 $\pm$ 0.1 & 0.925 &         190  &         0.060  &         6.655 $\pm$ 0.120  &         0.929 $\pm$ 0.017 \\
             &                 &      &                &       & 383 & 0.340 & 6.935 $\pm$ 0.120 & 0.968 $\pm$ 0.017 \\
9-ethynylphenanthrene & 7.84 $\pm$ 0.02 & 5.76 & 15.9 $\pm$ 0.1 & 1.013 &         320  &         0.290  &         7.337 $\pm$ 0.120  &         0.956 $\pm$ 0.016 \\
                      &                 &      &                &       & 413 & 0.513 & 7.560 $\pm$ 0.120 & 0.985 $\pm$ 0.016 \\
pyrene & 7.415 $\pm$ 0.010\tablenotemark{b} & 6.185 & 16.3 $\pm$ 0.1 & 1.013 &         320  &         0.241  &         8.112 $\pm$ 0.110  &         1.057 $\pm$ 0.015 \\
       &                                    &       &                &       & 493 & 0.672 & 8.544 $\pm$ 0.110 & 1.114 $\pm$ 0.015 \\
1-ethynylpyrene & 7.41 $\pm$ 0.02 & 6.19 & 16.4 $\pm$ 0.1 & 1.101 &         190  &         0.089  &         7.978 $\pm$ 0.120  &         0.975 $\pm$ 0.015  \\
                &                 &      &                &       & 403 & 0.525 & 8.414 $\pm$ 0.120 & 1.028 $\pm$ 0.015 \\
\enddata
\tablecomments{$E_{\mathrm{i}}$: ionization energy; $E^{\ast} = 13.6 - E_{\mathrm{i}}$: maximum internal energy of the parent cation in H~I regions; $E_{\mathrm{a},-\mathrm{H}}$: appearance energy of the singly dehydrogenated fragment ion; $\Delta E$: kinetic shift; $T$: vibrational temperature; $\left\langle E_{\mathrm{t}} \right\rangle$: average thermal energy; $E_{\mathrm{c},-\mathrm{H}}$: critical energy for the loss of a H atom by the parent cation; and $R$: photostability index.}
\tablenotetext{a}{Two values are given for each species: the translational temperature and the oven temperature, which are respectively considered as the lowest and highest values possibly taken by the vibrational temperature.}
\tablenotetext{b}{\cite{Mayer11}.}
\end{deluxetable}


\begin{thebibliography}{}
\bibitem[Allain et al.(1996a)]{Allain96a} Allain, T., Leach, S., \& Sedlmayr, E. 1996, \aap, 305, 602
\bibitem[Allain et al.(1996b)]{Allain96b} Allain, T., Leach, S., \& Sedlmayr, E. 1996, \aap, 305, 616
\bibitem[Allamandola et al.(1985)]{Allamandola85} Allamandola, L. J., Tielens, A. G. G. M., \& Barker, J. R. 1985, \apj, 290, L25
\bibitem[Bhatt \& Cami(2015)]{Bhatt15} Bhatt, N. H. \& Cami, J. 2015, \apjs, 216, 22
\bibitem[Beuther et al.(2014)]{Beuther14} Beuther, H., Ragan, S. E., Ossenkopf, V., et al. 2014, \aap, 571, A53
\bibitem[Beuther et al.(2008)]{Beuther08} Beuther, H., Semenov, D., Henning, Th., \& Linz, H. 2008, \apj, 675, L33
\bibitem[Boschi et al.(1972)]{Boschi72b} Boschi, R., Murrell, J. N., \& Schmidt, W. 1972, Faraday Discuss. Chem. Soc., 54, 116
\bibitem[Boschi \& Schmidt(1972)]{Boschi72a} Boschi, R. \& Schmidt, W. 1972, Tetrahedron Lett., 13, 2577
\bibitem[Brand \& Baer(1983)]{Brand83} Brand, W. A. \& Baer, T. 1983, Int. J. Mass Spectrometry Ion Phys., 49, 103
\bibitem[Canosa et al.(1997)]{Canosa97} Canosa, A., Sims, I. R., Travers, D., Smith, I. W. M., \& Rowe, B. R. 1997, \aap, 323, 644
\bibitem[Chastaing et al.(2001)]{Chastaing01} Chastaing, D., Le Picard, S. D., Sims, I. R., \& Smith I. W. M. 2001, \aap, 365, 241
\bibitem[Cherchneff \& Cau(1999)]{Cherchneff99} Cherchneff, I. \& Cau, P. 1999, in IAU Symp. 191, Asymptotic Giant Branch Stars, ed. T. Le Bertre, A. L{\`e}bre, \& C. Waelkens (San Francisco, CA: ASP), 251
\bibitem[Cherchneff et al.(1992)]{Cherchneff92} Cherchneff, I., Barker, J. R., \& Tielens, A. G. G. M. 1992, \apj, 401, 269
\bibitem[Chiar et al.(2013)]{Chiar13} Chiar, J. E., Tielens, A. G. G. M., Adamson, A. J., \& Ricca, A. 2013, \apj, 770, 78
\bibitem[Clayton et al.(2003)]{Clayton03} Clayton, G. C., Gordon, K. D., Salama, F., et al. 2003, \apj, 592, 947
\bibitem[Danks et al.(1984)]{Danks84} Danks, A. C., Federman, S. R., \& Lambert, D. L. 1984, \aap, 130, 62
\bibitem[Ervin et al.(1990)]{Ervin90} Ervin, K. M., Gronert, S., Barlow, S. E., et al. 1990, J. Am. Chem. Soc., 112, 5750
\bibitem[Frenklach et al.(1985)]{Frenklach85} Frenklach, M., Clary, D. W., Gardiner, W. C., \& Stein, S. E. 1985, Symposium (International) on Combustion, 20, 887
\bibitem[Frenklach \& Feigelson(1989)]{Frenklach89} Frenklach, M. \& Feigelson, E. D. 1989, \apj, 341, 372
\bibitem[Frisch et al.(2009)]{Gaussian09} Frisch, M. J., Trucks, G. W., Schlegel, H. B., et al. 2009, Gaussian 09, Revision A.02 (Wallingford CT: Gaussian, Inc.)
\bibitem[Garcia et al.(2013)]{Garcia13} Garcia, G. A., Cunha De Miranda, B. K., Tia, M., Daly, S., \& Nahon, L. 2013, Rev. Sci. Instrum., 84, 053112; erratum, Rev. Sci. Instrum., 84, 069902
\bibitem[Garcia et al.(2004)]{Garcia04} Garcia, G. A., Nahon, L., \& Powis, I. 2004, Rev. Sci. Instrum., 75, 4989
\bibitem[Gerin et al.(2010)]{Gerin10} Gerin, M., De Luca, M., Goicoechea, J. R., et al. 2010, \aap, 521, L16
\bibitem[Gerin et al.(2011)]{Gerin11} Gerin, M., Ka{\'z}mierczak, M., Jastrzebska, M., et al. 2011, \aap, 525, A116
\bibitem[Gotkis \& Lifshitz(1993)]{Gotkis93a} Gotkis, I. \& Lifshitz, C. 1993, Org. Mass Spectrom., 28, 372
\bibitem[Gotkis et al.(1993)]{Gotkis93b} Gotkis, Y., Oleinikova, M., Naor, M., \& Lifshitz, C. 1993, J. Phys. Chem., 97, 12282
\bibitem[Goulay \& Leone(2006)]{Goulay06a} Goulay, F. \& Leone, S. R. 2006, J. Phys. Chem. A, 110, 1875
\bibitem[Goulay et al.(2006)]{Goulay06b} Goulay, F., Rebrion-Rowe, C., Biennier, L., et al. 2006, J. Phys. Chem. A, 110, 3132
\bibitem[Gredel et al.(2011)]{Gredel11} Gredel, R., Carpentier, Y., Rouill{\'e}, G., et al. 2011, \aap, 530, A26
\bibitem[Hager \& Wallace(1988)]{Hager88} Hager, J. W. \& Wallace, S. C. 1988, Anal. Chem., 60, 5
\bibitem[Hahndorf et al.(2002)]{Hahndorf02} Hahndorf, I., Lee, Y. T., Kaiser, R. I., et al. 2002, J. Chem. Phys., 116, 3248
\bibitem[Henning et al.(2010)]{Henning10} Henning, Th., Semenov, D., Guilloteau, St., et al. 2010, \apj, 714, 1511
\bibitem[Herbig(1995)]{Herbig95} Herbig, G. H. 1995, \araa, 33, 19
\bibitem[Ho et al.(1995)]{Ho95} Ho, Y.-P., Dunbar, R. C., \& Lifshitz, C. 1995, J. Am. Chem. Soc., 117, 6504
\bibitem[Hudgins \& Sandford(1998)]{Hudgins98} Hudgins, D. M. \& Sandford, S. A. 1998, J. Phys. Chem. A, 102, 329
\bibitem[Huggins et al.(1984)]{Huggins84} Huggins, P. J., Carlson, W. J., \& Kinney, A. L. 1984, \aap, 133, 347
\bibitem[Innocenti et al.(2013)]{Innocenti13} Innocenti, F., Eypper, M., Stranges, S., et al. 2013, J. Phys. B: At. Mol. Opt. Phys., 46, 045002
\bibitem[J{\"a}ger et al.(2009)]{Jaeger09} J{\"a}ger, C., Huisken, F., Mutschke, H., Llamas-Jansa, I., \& Henning, Th. 2009, \apj, 696, 706
\bibitem[J{\"a}ger et al.(2007)]{Jaeger07} J{\"a}ger, C., Huisken, F., Mutschke, H., et al. 2007, Carbon, 45, 2981
\bibitem[Jenkins \& Tripp(2011)]{Jenkins11} Jenkins, E. B. \& Tripp, T. M. 2011, \apj, 734, 65
\bibitem[Jochims et al.(1999)]{Jochims99} Jochims, H. W., Baumg{\"a}rtel, H., \& Leach, S. 1999, \apj, 512, 500
\bibitem[Jochims et al.(1994)]{Jochims94} Jochims, H. W., R{\"u}hl, E., Baumg{\"a}rtel, H., Tobita, S., \& Leach, S. 1994, \apj, 420, 307
\bibitem[Jochims et al.(1997)]{Jochims97} Jochims, H. W., R{\"u}hl, E., Baumg{\"a}rtel, H., Tobita, S., \& Leach, S. 1997, Int. J. Mass Spectrom. Ion Process., 167/168, 35
\bibitem[Jones et al.(2010)]{Jones10} Jones, B., Zhang, F., Maksyutenko, P., Mebel, A. M., \& Kaiser, R. I. 2010, J. Phys. Chem. A., 114, 5256
\bibitem[Jones et al.(2011)]{Jones11} Jones, B. M., Zhang, F., Kaiser, R. I., et al. 2011, Proc. Natl. Acad. Sci. U.S.A., 108, 452
\bibitem[Kaiser et al.(1996)]{Kaiser96} Kaiser, R. I., Ochsenfeld, C., Head-Gordon, M., Lee, Y. T., \& Suits, A. G. 1996, Science, 274, 1508
\bibitem[Kassel(1928)]{Kassel28} Kassel, L. S. 1928, J. Phys. Chem., 32, 225
\bibitem[Khan(1992)]{Khan92} Khan, Z. H. 1992, Acta Phys. Pol. A, 82, 937
\bibitem[Larson(1981)]{Larson81} Larson, R. B. 1981, MNRAS, 194, 809
\bibitem[Lee et al.(2000)]{Lee00} Lee, S., Samuels, D. A., Hoobler, R. J., \& Leone, S. R. 2000, J. Geophys. Res., 105, 15085
\bibitem[Leger \& Puget(1984)]{Leger84} Leger, A. \& Puget, J. L. 1984, \aap, 137, L5
\bibitem[Le Page et al.(2001)]{LePage01} Le Page, V., Snow, T. P., \& Bierbaum, V. M. 2001, \apjs, 132, 233
\bibitem[Le Page et al.(2003)]{LePage03} Le Page, V., Snow, T. P., \& Bierbaum, V. M. 2003, \apj, 584, 316
\bibitem[Lifshitz(1991)]{Lifshitz91} Lifshitz, C. 1991, Int. J. Mass Spectrom. Ion Process., 106, 159
\bibitem[Ling et al.(1995)]{Ling95} Ling, Y., Gotkis, Y., \& Lifshitz, C. 1995, Eur. J. Mass Spectrom., 1, 41
\bibitem[Ling \& Lifshitz(1998)]{Ling98} Ling, Y. \& Lifshitz, C. 1998, J. Phys. Chem. A, 102, 708
\bibitem[Liszt et al.(2012)]{Liszt12} Liszt, H., Sonnentrucker, P., Cordiner, M., \& Gerin, M. 2012, \apj, 753, L28
\bibitem[Lucas \& Liszt(2000)]{Lucas00} Lucas, R. \& Liszt, H. S. 2000, \aap, 358, 1069
\bibitem[Maksyutenko et al.(2011)]{Maksyutenko11} Maksyutenko, P., Zhang, F., Gu, X., \& Kaiser, R. I. 2011, Phys. Chem. Chem. Phys., 13, 240
\bibitem[Mayer et al.(2011)]{Mayer11} Mayer, P. M., Blanchet, V., \& Joblin, C. 2011, J. Chem. Phys., 134, 244312
\bibitem[Mebel et al.(2008)]{Mebel08} Mebel, A. M., Kislov, V. V., \& Kaiser, R. I. 2008, J. Am. Chem. Soc., 130, 13618
\bibitem[Mercier et al.(2000)]{Mercier00} Mercier, B., Compin, M., Prevost, C., et al. 2000, J. Vac. Sci. Technol. A, 18, 2533
\bibitem[Montillaud et al.(2013)]{Montillaud13} Montillaud, J., Joblin, C., \& Toublanc, D. 2013, \aap, 552, A15
\bibitem[Nahon et al.(2012)]{Nahon12} Nahon, L., De Oliveira, N., Garcia, G. A., et al. 2012, J. Synchrotron Rad., 19, 508
\bibitem[Ohishi et al.(1992)]{Ohishi92} Ohishi, M., Irvine, W. M., \& Kaifu, N. 1992, in IAU Symp. 150, Astrochemistry of Cosmic Phenomena, ed. P. D. Singh (Dordrecht: Kluwer), 171
\bibitem[Parisel et al.(1992)]{Parisel92} Parisel, O., Berthier, G., \& Ellinger, Y. 1992, \aap, 266, L1
\bibitem[Phillips \& Huggins(1981)]{Phillips81} Phillips, T. G. \& Huggins, P. J. 1981, \apj, 251, 533
\bibitem[Reed \& Kass(2000)]{Reed00} Reed, D. R. \& Kass, S. R. 2000, J. Mass Spectrom., 35, 534
\bibitem[Richard-Viard et al.(1996)]{RichardViard96} Richard-Viard, M., Delboulb{\'e}, A., \& Vervloet, M. 1996, Chem. Phys., 209, 159
\bibitem[Rice \& Ramsperger(1927)]{Rice27} Rice, O. K. \& Ramsperger, H. C. 1927, J. Am. Chem. Soc., 49, 1617
\bibitem[Rouill{\'e} et al.(2012)]{Rouille12} Rouill{\'e}, G., Steglich, M., Carpentier, Y., et al. 2012, \apj, 752, 25
\bibitem[Rouill{\'e} et al.(2013)]{Rouille13} Rouill{\'e}, G., J{\"a}ger, C., Huisken, F., \& Henning, Th. 2013, Proc. Int. Astron. Union 9 (S297), 276
\bibitem[Salama et al.(2011)]{Salama11} Salama, F., Galazutdinov, G. A., Kre{\l}owski, J., et al. 2011, \apj, 728, 154
\bibitem[Schilke et al.(1995)]{Schilke95} Schilke, P., Keene, J., Le Bourlot, J., Pineau des For{\^e}ts, G., \& Roueff, E. 1995, \aap, 294, L17
\bibitem[Shi \& Ervin(2000)]{Shi00} Shi, Y. \& Ervin, K. M. 2000, Chem. Phys. Lett., 318, 149
\bibitem[Stark et al.(1996)]{Stark96} Stark, R., Wesselius, P. R., van Dishoeck, E. F., \& Laureijs, R. J. 1996, \aap, 311, 282
\bibitem[Steglich et al.(2011)]{Steglich11} Steglich, M., Bouwman, J., Huisken, F., \& Henning, Th. 2011, \apj, 742, 2
\bibitem[Steglich et al.(2012)]{Steglich12} Steglich, M., Carpentier, Y., Jäger, C., et al. 2012, \aap, 540, A110
\bibitem[Thantu \& Weber(1993)]{Thantu93} Thantu, N. \& Weber, P. M. 1993, Z. Phys. D, 28, 191
\bibitem[Tucker et al.(1974)]{Tucker74} Tucker, K. D., Kutner, M. L., \& Thaddeus, P. 1974, \apj, 193, L115
\bibitem[Turner et al.(2000)]{Turner00} Turner, B. E., Herbst, E., \& Terzieva, R. 2000, \apjs, 126, 427
\bibitem[van der Meij et al.(1988)]{vanderMeij88} van der Meij, C. E., van Eck, J., \& Niehaus, A. 1988, Chem. Phys., 119, 135
\bibitem[Wang \& Frenklach(1994)]{Wang94} Wang, H. \& Frenklach, M. 1994, J. Phys. Chem., 98, 11465
\bibitem[West et al.(2014a)]{West14a} West, B., Sit, A., Bodi, A., Hemberger, P., \& Mayer, P. M. 2014a, J. Phys. Chem. A, 118, 11226
\bibitem[West et al.(2014b)]{West14b} West, B., Useli-Bacchitta, F., Sabbah, H., et al. 2014b, J. Phys. Chem. A, 118, 7824
\bibitem[Wootten et al.(1980)]{Wootten80} Wootten, A., Bozyan, E. P., Garrett, D. B., Loren, R. B., \& Snell, R. L. 1980, \apj, 239, 844
\bibitem[Zhang et al.(2010)]{Zhang10} Zhang, J., Han, F., \& Kong, W. 2010, J. Phys. Chem. A, 114, 11117
\bibitem[Zhen et al.(2015)]{Zhen15} Zhen, J., Castellanos, P., Paardekooper, D. M., et al. 2015, \apj, 804, L7
\end{thebibliography}
\end{document}